\documentclass[amssymb,prb,preprint,aps,twocolumn,showpacs,11pt]{revtex4}
\usepackage{epsfig}
\usepackage{mathrsfs}
\begin{document}

\title{Finite width of the sonic event horizon and grey body Hawking radiation.}
\author{Y. Vinish V. Fleurov}
\affiliation{Raymond and Beverly Sackler Faculty of Exact Sciences,\\
School of Physics and Astronomy, Tel-Aviv University, Tel-Aviv
69978, Israel
}

\begin{abstract}
Finite width of the analog event horizon is determined by the nonlinearity length in the Kerr nonlinear optical system, which is discussed here, or by the healing length in Bose-Einstein condensates. The various eigen modes of fluctuations are found in the immediate vicinity of the event horizon and the scattering matrix due to the finite width horizon is calculated to within the leading order corrections in the nonlinearity length. The Hawking radiation is found to be that of a grey body with the emissivity larger than one. A procedure of paraxial quantization of the fluctuation field is discussed and its connection to the conventional quantization of the electromagnetic field is demonstrated.
\end{abstract}

 \maketitle

\section{Introduction}

Analogue gravity modelled in various laboratory systems is a rapidly developing field both from theoretical and experimental points of view. The seminal paper by Unruh \cite{U81} proposed a transonic flow of a barotropic isentropic fluid as a simulator of a black hole event horizon. The prediction was that a radiation analogous to the celebrated Hawking radiation \cite{H75,H76} could be observed in such a system. Since then a number of various systems were proposed as playgrounds for simulating the event horizon of black hole.\cite{JV98,R00,G05,BLV03,CFRBF08,RPC09,NBRB09,FFBF10,BCP14}
There is also an important progress achieved in experiment, e.g.  a white-hole horizon, created by a moving soliton, was observed in optical fibers.\cite{PKRHK08} An observation of analog Hawking radiation in optical fibers was reported.\cite{BCCGORRSF10} A black-hole horizon was observed in a Bose-Einstein condensate (BEC) system,\cite{LIBGRZS10} and quite recently an observation of laser type amplification of Hawking radiation\cite{Jeff14} has been reported. A "horizon physics" is studied also in the surface water waves.\cite{RMMPL08,RMMPL10,WTPUL11} A possibility of creating an event horizon in the coherent light propagating in the Kerr nonlinear defocusing medium was discussed in Ref. \cite{FFBF10,EFB12,EBFS13} Reviews on the progress in the field and relevant topics can be found in recent papers.\cite{R12,C13,BF14}

One of the intriguing questions, which is common for general relativity (GR) black holes and analogue gravity models, is the behavior of the radiation in the immediate vicinity of the event horizon. The wave length becomes of the order of Planck length in GR and healing length in BEC or nonlinearity length in the optical analogues of the horizon. The issue was addressed in Refs. \onlinecite{J91,J93,U95}. Quantitative results in GR were obtained by introducing a sub- or superluminal deviation of the otherwise linear spectrum of massless particles, which may happen on the Planckian (healing length) scale, \cite{C97,MP09a,MP09b,RPC09,BMPS95,CJ96,C98,CJ99,LKO03a,LKO03b,US05,SU08,FP11} A review can be found in Ref. \onlinecite{BLV05}

Analysis of the role of the quantum potential in Ref. \onlinecite{FS12} shows that the behavior of fluctuations becomes regular near the event horizon on the scale $l_r$, which is somewhat larger than the healing length. A similar regularization length appears in the numerical study,\cite{FP11} (see also discussion in Ref. \onlinecite{CP14} of the various length scales appearing in the problem in the GR context).

The approach outlined in Ref. \onlinecite{FS12} will allow us to find explicit formulas for all eigenmodes for fluctuations near the event horizon. Two of these eigenmodes are evanescent in the subsonic flow but become real after a certain critical distance from the horizon in the supersonic region of the flow.

Considering Hawking radiation in an all-optical configuration \cite{FFBF10,EFB12,EBFS13} we have to address several issues, some of which are also relevant to other experimental setups. The most important one is the finite width of the event horizon, due to the processes taking place on the scale of the nonlinearity length. (A finite width of the event horizon was recently discussed in Ref. \onlinecite{CP14} within the GR context. The uncertainty in the position of the Schwarzschild sphere due to noncommutation of the two metric components was considered in Ref. \onlinecite{DY14}). They determine formation of the six types of fluctuation --- two positive frequency modes in the subsonic region and four positive and negative frequency modes in the supersonic region. Calculation of the scattering matrix and spectrum of the Hawking radiation for finite values of the nonlinearity length becomes then straightforward. (Scattering problem in BEC was considered for several specific potential and interaction coefficient shapes. \cite{RPC09,LRCP12,LPK12})

Another important issue is the quantization of fluctuations. Analysis of the analogue Hawking radiation in an all-optical setup is based on the
nonlinear Schr\"odinger equation (NLS)
\begin{equation}\label{NLS}
i \partial_z A =
-\frac{1}{2 \beta_0}\widetilde{\nabla}^2A + g |A|^2 A
\end{equation}
deduced from the classical Maxwell equations in the paraxial approximation. Here $A$ is the amplitude of the electric field, the wave vector $\beta_0$ of the light plays the role analogous to the mass of a "quantum particle". The propagation distance $z$ is now "time". The Laplacian now contains derivatives with respect to the coordinates $x$, $y$ and real time $t$.
Since the part of time is played now by the propagation distance $z$ the quantization in this approximation, i.e. introduction of "paraxial photons" becomes rather tricky. A general approach to the paraxial quantization was proposed in Refs. \onlinecite{DG91,AW05,AML10}, which allows one to connect the standard quantization of the electromagnetic field to the paraxial photons. Paraxial quantization of the fluctuations near the all-optical event horizon will be carried out below.

\section{Fluctuations near the event horizon in a luminous fluid}

The propagation of coherent light in a Kerr nonlinear medium in the paraxial approximation can be mapped on a flow of an equivalent luminous fluid. Madelung transformation $A= f e^{-i\varphi}$ allows one to represent the NLS equation (\ref{NLS}) in the form of two hydrodynamic equations
for the density $\rho({\bf r},z) = \beta_0 f^2({\bf r},z)$, which is in fact the light intensity, and velocity ${\bf v}({\bf r},z) = - \frac{1}{\beta_0}\nabla \varphi({\bf r},z) $.

We will consider here small fluctuations of the amplitude $\delta A = A - A_0$ with respect to a stationary solution $A_0 = f_0 e^{-i\varphi_0}$. Their dynamics is described by the equations
\begin{eqnarray}
  \widehat{D} \chi - \displaystyle \frac{1}{\beta_0}\frac{1}{f_0^2}
\nabla(f_0^2 \nabla \xi) &=& 0 \label{linear-1}\\
  \displaystyle \widehat{D} \xi + \frac{1}{4\beta_0}
\frac{1}{f_0^2}\nabla( f_0^2 \nabla \chi) - gf_0^2
\chi &=& 0 \label{linear-2}
\end{eqnarray}
obtained by linearizing Eq. (\ref{NLS}) (see, e.g. Refs. \cite{FS11,FS12}).
Here $\widehat{D} = \partial_z + {\bf v}_0 \cdot \nabla $ and
\begin{equation}\label{operators}
\begin{array}{c}
\chi = \frac{1}{f_0} \left[ e^{-i\varphi_0} \delta A^* +
e^{i\varphi_0} \delta A \right],
\\
\xi = \frac{1}{2if_0}[e^{-i\varphi_0} \delta A^* - e^{i\varphi_0}
\delta A],
\end{array}
\end{equation}
are classical scalar fields describing fluctuations of the amplitude and phase, respectively. Although the functions $\chi$ and $\xi$ are explicitly real, we will consider below the general properties of the complex solutions of the linear equations (\ref{linear-1}) and (\ref{linear-2}). However when calculating the physically measurable quantities only the real part of the functions should be considered.

We now assume that the stationary solution behaves as
$\rho_0 = \beta_0 f^2_0(x) = \frac{\beta_0\overline{s}^2}{g}(1 - \alpha x)$ and $v_0(x) = \overline{s} (1 + \alpha x)$ with a parameter $\alpha$. Here $\overline{s}$ is the sound velocity of the luminous fluid at $x=0$, and $x$ is the distance from the horizon surface along the streamline normal to it. This approximation holds at $\alpha x \ll 1$.

It is sufficient to limit the discussion to $1+1$ dimensions. Then following the derivation outlined in Refs. \onlinecite{FS12,EBFS13} we get the solutions of the equations (\ref{linear-1}) and (\ref{linear-2}) as integrals with the properly chosen integration contours:
\begin{widetext}
\begin{equation}\label{integral}
\chi(x,z) = \int d\omega e^{- i\omega z} \int_C dk k ^{\gamma_1} \left( k -
\frac{2}{3} \nu - \frac{i}{3}\alpha \right)^{\gamma_2} \exp\left\{
\Lambda(k,\nu) + ikx\right\}.
\end{equation}
\end{widetext}
where
$$
\gamma_1 = \frac{1}{4} - \frac{i \nu}{2\alpha},
$$$$
\gamma_2 = -
\frac{1}{4} - i \frac{1}{6\alpha}\nu  - \frac{4i}{81 \alpha} l_n^2
\nu^3 + \frac{14}{81} l_n^2 \nu^2
$$
and the $l_n$ dependent part is given by
\begin{widetext}
\begin{equation}\label{solution-b}
\Lambda(k,\nu) = \frac{l_n^2}{\alpha} \left \{- \frac{i }{18} k^3 +
\frac{5}{36} \alpha k^2  - \frac{i}{18} \nu k^2 - \frac{2i}{27}
\nu^2 k + \frac{4}{27} \nu \alpha k \right \}.
\end{equation}
\end{widetext}
$l_n^2 = \frac{1}{2\beta_0^2 \overline{s}^2}$ is the nonlinearity length and $\nu = \omega/\overline{s}$ is the "frequency" scaled with the sound velocity.

Integral (\ref{integral}) can be also calculated by means of the steepest descent technique. For this we first have to find the saddle points determined by the equation
\begin{equation}\label{difequ}
\nu - k v(x) = \pm \Omega(k,\nu, x)
\end{equation}
where
\begin{equation}\label{saddle-points}
\Omega^2(k,\nu,x) =\displaystyle\frac{l_n^2}{2} (i\alpha k + k^2)^2 \overline{s}^2 + k^2 s^2(x)
\end{equation}
obtained in Ref. \onlinecite{FS12}.

Neglecting the small $\alpha/ k \ll 1$ corrections in the quartic term, we have an equation that looks exactly as the Bogolubov dispersion relation for the above condensate excitations in the moving frame. The important difference, however, is that both the sound velocity and the flow velocity depend on the coordinate and the solutions may change drastically when crossing the event horizon (at $x = 0$). The conventional Bogolubov spectrum of excitations in NLS equation is obtained under the condition that the amplitude $f_0$ and velocity $v_0$ are constants. In the context of our problem these conditions may be fulfilled at large distances from the sonic horizon, whereas Eq. (\ref{difequ}) holds in its immediate vicinity. It allows one to follow the evolution and interconnection of the eigen functions, when passing from the subsonic region ($x < l_r$) via regularization region $|x| < l_r$ to the supersonic region $x > l_r$. Here $l_r=l_n/(\alpha l_n)^{1/3}$ is the regularization length.
\begin{figure}[tbp!]
\begin{center}
\includegraphics[width=1\columnwidth]{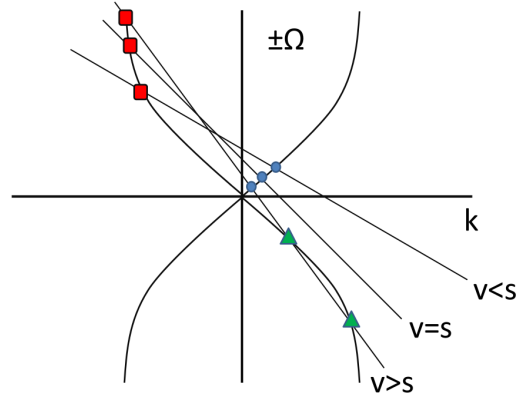}
\end{center}
\caption{(Color online) Graphical analysis of equation
(\ref{difequ}). Two branches of the function $\Omega(k,\omega)$
are plotted. They are crossed by the
straight dashed lines $\nu - kv(x)$ for three positions of an "observer": $x < 0$ ---
subsonic regime, when $v < s$, $x = 0$, at the horizon, when $v = s$; and $x > 0$ ---
supersonic regime, when $v > s$.} \label{saddle.fig}
\end{figure}

Graphical solutions of Eq. (\ref{difequ}) are shown in Figure \ref{saddle.fig}. In principle, $\Omega(k,\nu, x)$ also varies with $x$ but we ignore it in the graph in order not to overload it. In the analytical calculations this dependence is certainly taken into account.
Three blue circles show the solution with the saddle point at $k_r \approx \nu/2\overline{s}$,  which we call regular. It is obtained under the assumption that $k_rl_n \ll 1$ and corresponds to the fluctuation $\chi_r = e^{-i\nu \left(z - \frac{x}{2\overline{s}}\right)}$ propagating downstream with the double sound velocity. This solution changes only slightly when moving from subsonic to supersonic region.

The red squares show the evolution of the solution corresponding to the saddle point $k_s = 2\nu/3\overline{s}\alpha x$ in the subsonic region ($x < -l_r$) also obtained under the condition $k_sl_n\ll 1$, i.e. $|x|/l_n\gg 2\nu/3\overline{s}\alpha$. When moving closer to the horizon and crossing it (see figure \ref{saddle.fig}) this solution moves towards large $k$ values where the limit $kl_n \gg 1$ should be taken. Then equation (\ref{difequ}) becomes
\begin{equation}\label{saddle-points-b}
\frac{l_n^2}{2}  k^3 - 3 k \alpha x + 2\nu = 0
\end{equation}
This equation has three solutions one of which is real and the other two become real only at
$$
x \geq x_c = \left(\frac{l_n^2\nu^2}{2\alpha^3}\right)^{1/3} = \frac{l_r}{2^{1/3}} \left(\frac{\nu}{\alpha}\right),
$$
with the corresponding wave vector $k_c = \left(\frac{2\nu}{l_n^2}\right)^{1/3}$. These two modes appear due to bifurcation in the lower half plane in Fig. \ref{saddle.fig} when the straight line representing the l.h.s. of Eq. (\ref{difequ}) touches the curve $-\Omega(k,\nu, x)$. This bifurcation point coincides to within a numerical factor with the turning point found in Ref. \onlinecite{CPF14}.

The real solution of (\ref{saddle-points-b}) is $k_{1h}= (4\nu/l_n^2)^{1/3}$ within the width of the horizon at $|x| \ll x_c$ and $k_{e1}= (6\alpha x)^{1/2}/l_n$ outside at $x \gg x_c$. The corresponding eigenfuntion describes a mode, which propagates upstream. That is why its character changes drastically when moving from the subsonic to supersonic region, from the singular function $\chi_{s1} = e^{-i\nu z} x^{\gamma -1}$ to $\chi_{1h} = e^{- i\nu z - i(4\nu/l_n^2)^{1/3} x} $ at $|x| \ll x_c$ to $\chi_{e1} = e^{- i\nu z - i \frac{2\sqrt{6\alpha}}{3l_n}x^{3/2}}$. Here $\gamma = - \gamma_1 - \gamma_2.$

The eigenfunction $\chi_{e1}$ appears only due to the quartic term in the fluctuation spectrum in Eq. (\ref{difequ}) and is related to the evanescent solution in the subsonic region. The other two eigenfunctions appear due to bifurcation in the lower half plane in Fig. \ref{saddle.fig}, which takes place at $x = x_c$. The two emerging saddle points, one moving towards smaller $k$ ($k^3$ can be neglected in (\ref{saddle-points-b})) and the other one towards large $k$ (free term can be neglected) produce two eigenfunctions: the singular $\chi_{s2} = e^{-i\nu z} x^{\gamma -1}$ and the one related to the second evanescent function $\chi_{e2} = e^{- i\nu z + i \frac{2\sqrt{6\alpha}}{3l_n}x^{3/2}}$. Both functions $\chi_{e1}$ and $\chi_{e2}$ exist as propagating waves only in the supersonic region, otherwise they become evanescent (see also discussion in Ref. \onlinecite{LRCP12}).

\begin{figure}[tbp!]
\begin{center}
\includegraphics[width=1.3\columnwidth]{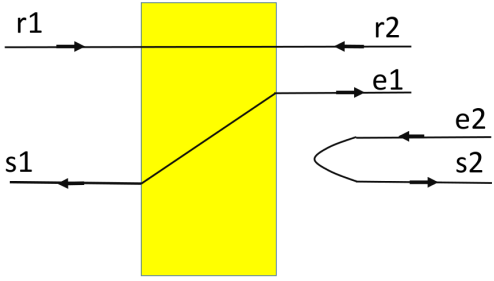}
\end{center}
\caption{(Color online) The figure shows schematically evolution of solutions of Eq. (\ref{difequ}) when moving from the subsonic region through the event horizon (yellow box) to the supersonic region.} \label{solutions.fig}
\end{figure}

\section{Scattering matrix.}

Now we are in a position to calculate the scattering matrix for an event horizon of a small but finite width. First we have to make use of the coordinates
\begin{eqnarray}
   x &\rightarrow& \tilde{x}=x \label{x-trans}\\
   z &\rightarrow& z+\int\frac{v_0(x)dx}{s^2(x)-v^2_0(x)}\approx z-\frac{\ln(x)}{3\alpha\bar s} \label{z-trans}
 \end{eqnarray}
This coordinate transformation was used in Ref. \onlinecite{U81} in order to diagonalize the metric and represent it in the conventional Schwarzschild form. Then the density and current flow corresponding to the canonical pair of fields $\chi$ and $\xi$ take the form\cite{FFBF10}
\begin{eqnarray}
\varrho = -i\frac{s^2(\tilde x)}{s^2(\tilde x)-{v_0}^2(\tilde x)}[(\partial_{\tilde z}\xi^*)\xi-\xi^*(\partial_{\tilde z}\xi)] \label{density}\\
j = -i[{v_0}^2(\tilde x)-s^2(\tilde x)][(\partial_{\tilde x}\xi^*)\xi-\xi^*(\partial_{\tilde x}\xi)]
  \label{current density}
\end{eqnarray}
where the relation $\chi \approx \frac{1}{\bar s}\partial_x\xi$, holding outside the regularization region ($|x|\gg l_r$), has been used.

The scattering matrix $S$ transforms three incoming waves $(r1,r2,e2)$, of which the two last waves are negative frequency waves, into three outgoing waves $(s1,s2,e1)$. Hence the unitarity condition $S^\dag U S = U$ is defined with $U=\mbox{diag}(1,-1,-1)$. The balance of the incoming and outgoing currents
\begin{equation}\label{currents}
\begin{array}{c}
  j_{s1} = |S_{11}|^2j_{r1} -|S_{12}|^2 j_{r2} - |S_{13}|^2j_{e2}, \\
  j_{s2} = |S_{21}|^2j_{r1} -|S_{22}|^2 j_{r2} - |S_{23}|^2j_{e2}, \\
  i_{e1} = |S_{31}|^2j_{r1} -|S_{32}|^2 j_{r2} - |S_{33}|^2j_{e2},
\end{array}
\end{equation}
holds under the condition that the incoming functions
\begin{equation}\label{incoming}
\begin{array}{c}
  \xi_{r1}(x) =  |\tilde x|^{-\frac{\gamma_0}{2}} e^{-i\nu\tilde z}, \\
  \xi_{r2}(x) =  \tilde x^{-\frac{\gamma_0}{2}} e^{-i\nu\tilde z}, \\
  \xi_{e2}(x) = \sqrt{\frac{2l_n\nu}{(6\alpha\bar x)^{3/2}}} \tilde x^{-\frac{\gamma_0}{2}}e^{-i\nu\tilde z + i \frac{\sqrt{2\alpha}}{3l_n}x^{3/2}},
\end{array}
\end{equation}
and outgoing functions
\begin{equation}\label{outgoing}
    \begin{array}{c}
      \xi_{s1}(x) = |\tilde x|^{\frac{\gamma_0}{2}} e^{-i\nu\tilde z}, \\
      \xi_{s2}(x) = \tilde x^{\frac{\gamma_0}{2}} e^{-i\nu\tilde z}, \\
      \xi_{e1}(x) = \sqrt{\frac{2l_n\nu}{(6\alpha\bar x)^{3/2}}} \tilde x^{-\frac{\gamma_0}{2}} e^{-i\nu\tilde z - i \frac{2\sqrt{6\alpha}}{3l_n}x^{3/2}},
    \end{array}
\end{equation}
are properly normalized. Here $\bar x$ gives us a scale where the linear $x$ dependence of the flow velocity holds, $\bar x \approx 1/\alpha$. It corresponds to one of the length scales discussed in Ref. \onlinecite{CP14}. The normalization cannot be carried out directly, since we know the eigen functions only in a limited part of the space and cannot integrate the density (\ref{density}) in the whole space. However, we can find a relative normalization. So that all the currents in (\ref{currents}) were equal. Then Eqs. (\ref{currents}) become compatible with the unitarity of the scattering matrix. This has allowed us to choose the coefficients in Eqs. (\ref{incoming}) and (\ref{outgoing}), so that this functions are now defined within a common factor.

Similar eigen modes are obtained in Refs. \onlinecite{CPF14,BP12} in the GR context. The mode equation in this case differs from our equations (\ref{linear-1}) and (\ref{linear-2}) and produce eigen modes, which differ from those obtained in the previous section. They become similar only after the transformation (\ref{x-trans}) and (\ref{z-trans}).

The scattering matrix has the simple form
\begin{equation}\label{l=0}
    S^{(0)}=\left(
      \begin{array}{ccc}
        \alpha_B & \beta_B & 0 \\
        \beta_B & \alpha_B & 0 \\
        0 & 0 & 1 \\
      \end{array}
    \right)
\end{equation}
in the limit $l_n \to 0$. Here the condition for the Bogolubov coefficients $\alpha_B^2 - \beta_B^2 =1$ follows from the unitarity of $S^{(0)}$ and $\alpha_B/\beta_B = e^{\pi{\rm Im}\gamma}$ results from the branch point, characteristic of the functions $\xi_{s1}$ and $\xi_{s2}$. This S-matrix approach is just another version of the analysis of Hawking radiation as presented in Refs. \onlinecite{DR76,DL08}.

Now we will calculate a matrix $S_{ij} = S^{(0)}_{ij} + S^{(1)}_{ij} + S^{(2)}_{ij}$ with the first and second order corrections due to small but nonzero $l_n$. The unitarity of this matrix results in six equations
\begin{equation}\label{currents-a}
\begin{array}{c}
  |S_{11}|^2 - |S_{12}|^2 - 1 = |S_{13}|^2 \\
  |S_{22}|^2 - |S_{12}|^2 - 1 =  - |S_{23}|^2  \\
  |S_{33}|^2 - 1 =  |S_{13}|^2  - |S_{23}|^2 \\
  S_{11}S_{12} - S_{21} S_{22} = S_{31}S_{32} \\
  S_{11}S_{13} - S_{21} S_{23} = S_{31}S_{33}\\
  S_{12}S_{13} - S_{22} S_{23} = S_{32}S_{33}
\end{array}
\end{equation}
The matrix elements $S_{13}$ and $S_{23}$ will be treated as small parameters. The matrix $S_{ij}$ will be assumed to be real in what follows. The last two equations in (\ref{currents-a}) in the leading order become
$$
 \begin{array}{c}
    (\alpha_B -1 )S_{13} = \beta_B  S_{23}  \\
   \beta_B S_{13} = (\alpha_B +1) S_{23}
 \end{array}
$$
and we get that
\begin{equation}\label{ratio-a}
\frac{S_{13}}{S_{23}} = e^{\pi{\rm Im}\gamma} + \sqrt{e^{2\pi{\rm Im}\gamma} -1} = h.
\end{equation}

The first order corrections in the fourth equation in (\ref{currents-a}) must compensate  each other, which yields the ratio
\begin{equation}\label{ratio-b}
\frac{ S^{(1)}_{11}}{S^{(1)}_{12}} = \frac{S^{(1)}_{22}}{S^{(1)}_{12}} = \frac{\beta_B}{\alpha_B} = e^{-\pi{\rm Im}\gamma}.
\end{equation}
It is also consistent with the first two equations.

The third equation in (\ref{currents-a}) is solved straightforwardly,
$
 S^{(1)}_{33} = 0,\ \ \ S^{(2)}_{33} = S^2_{23} \frac{h}{\beta}.
$
The remaining first, second and fourth equations, containing only second order corrections are linearly dependent and it is sufficient to consider only two of them,
\begin{equation}\label{currents-b}
\begin{array}{c}
  2\alpha S^{(2)}_{11} - 2\beta S^{(2)}_{12} = |S_{23}|^2 h^2 - {S^{(1)}_{11}}^2 + {S^{(1)}_{12}}^2  \\
  2\alpha S^{(2)}_{22} - 2\beta S^{(2)}_{12} = -|S_{23}|^2 - {S^{(1)}_{22}}^2 + {S^{(1)}_{12}}^2
\end{array}
\end{equation}

The determinant of the scattering matrix
$$
\begin{array}{c}
    \det S = 1 + (S^{(1)}_{11} S^{(1)}_{22} - S^{(1)}_{21} S^{(1)}_{12}) + \\
    S_{31}^2 (h^2 - 1) + S_{13}^2[2\beta_B h - \alpha_B (h^2 +1)]
  \end{array}
$$
is calculated using the above relations between the matrix elements. Requiring that the determinant equals to one and applying (\ref{ratio-b}) we get
$
{S^{(1)}_{11}} = {S^{(1)}_{22}} = {S^{(1)}_{12}}=0.
$

Now we have to go back to the fifth and sixth equations and consider the higher order terms
$$
\begin{array}{ccc}
  S^{(2)}_{11}S_{13} - S^{(2)}_{21} S_{23} & = & S_{31}S^{(2)}_{33} \\
  S^{(2)}_{12}S_{13} - S^{(2)}_{22} S_{23} & = & S_{32}S^{(2)}_{33}
\end{array}
$$
Then using the above relations between the matrix elements and Eqs. (\ref{currents-b}) we find all the second order correctiona.
%
As a result we get the scattering matrix
\begin{equation}\label{lneqzero}
    S =
$$$$
    \left(
      \begin{array}{ccc}
        \alpha_B + \frac{1}{2} h^2 S_{32}^2 & \beta_B + \frac{1}{2} h S_{32}^2& h S_{32} \\
        \beta_B + \frac{1}{2} h S_{32}^2 & \alpha_B + \frac{1}{2} S_{32}^2 & S_{32} \\
        h S_{32} & S_{32} & 1 + \frac{h}{\beta}  S_{32}^2 \\
      \end{array}
    \right)
\end{equation}

The fact that the functions $\xi_{s2}$ and $\xi_{e2}$ stem from the same bifurcation point where they must coincide, allows us to assume that $S_{32} \approx \sqrt{\frac{2l_n\nu}{(6\alpha\bar x)^{3/2}}} \propto \sqrt{l_n\nu} \sim l_n \sqrt{2\beta_0\omega}$, which is a rather crude estimate, holding to within a numerical factor. However, it may correctly reflect the dependence on $l_n$ which is of primary importance for us here.

We can now find the spectrum of the Hawking radiation (in the subsonic region)
$$
N_{H<}(\omega) = S_{12}^2 = \frac{g_<}{e^{\hbar\nu/T_H(\nu)} - 1}
$$
This is radiation of a grey body with the emissivity
\begin{equation}\label{grey}
g_< = 1 + l_n\nu h(\nu) \beta_B(\nu)
\end{equation}
with
\begin{eqnarray*}
  \beta_B(\nu) &=& \frac{1}{\sqrt{e^{\hbar\nu/T_H(\nu)} - 1}}, \\
  h(\nu) &=& e^{\hbar\nu/2T_H(\nu)} +\sqrt{e^{\hbar\nu/T_H(\nu)} - 1}
\end{eqnarray*}
and Hawking temperature
$$
T_H(\nu) = \frac{3 \hbar \overline{s} \alpha}{4\pi k_B} \left[1 + \frac{2 l_n^2 \nu^2 }{27\overline{s}^2}\right]^{-1}.
$$

It is important to emphasize a surprising result that the emissivity (\ref{grey}) is larger than one. It is well known that emissivity of any grey body at equilibrium with the photon gas is always smaller than one. Here, however we deal with an essentially nonequilibrium system. That is why the emissivity may be larger than one. It depends on the wave number $\nu$ of the emitted photon and increases with it.

The radiation in the supersonic region ("inside the black hole") occurs in two modes: (1) The negative frequency singular mode $s2$ radiates with the spectrum
$$
N_{H>}(\omega) = S_{22}^2 = \frac{g_{>}e^{\hbar\nu/T_H(\nu)}}{e^{\hbar\nu/T_H(\nu)}- 1}
$$
where $g_> = 1 + l_n \nu\sqrt{1 - e^{-\hbar\nu/T_H(\nu)}}$; (2) There is also a weak radiation $O(l_n\nu)$ due to the $e1$ mode.

\section{Paraxial quantization}

The analysis of Hawking radiation carried out above is based on the equations (\ref{linear-1}) and (\ref{linear-2}) deduced from the classical Maxwell equations in the paraxial approximation. However, the phenomenon of Hawking radiation is a quantum effect.  In this context quantization of the fluctuations becomes an important issue. As was shown in Ref. \onlinecite{FS11,EBFS13} equations (\ref{linear-1}) and (\ref{linear-2}) can be generated by the Lagrangian
\begin{equation}\label{Lagrangian-a}
{\cal L} = \frac{1}{2} f_0^2(\chi\partial_z\xi - \xi\partial_z\chi) + W(\chi,\xi)
\end{equation}
where
\begin{widetext}
\begin{equation}\label{W}
W(\chi,\xi) = \frac{1}{2} f_0^2 v_0 (\chi \partial_x \xi - \xi \partial_x \chi)
- \frac{1}{2}g f_0^4 \chi^2 - \frac{1}{2\beta_0} f_0^2
(\partial_x\xi)^2 - \frac{1}{8\beta_0} f_0^2 (\partial_x\chi)^2
\end{equation}
\end{widetext}

The quantization procedure seems to be straightforward.
Applying the Dirac procedure\cite{D64} we first get two constraints
\begin{equation}\label{canonicalmomentum}
\begin{array}{ccc}
\phi_\xi &=& p_\xi - \frac{1}{2} f_0^2 \chi = 0 \\
\phi_\chi &=& p_\chi + \frac{1}{2} f_0^2 \xi = 0
\end{array}
\end{equation}
connecting canonical momenta and coordinates at the classical solutions.
The Hamiltonian then becomes
$$
H = W(\chi,\xi) + \frac{1}{f_0^2}\left[\frac{\delta W(\chi,\xi)}{\delta \chi}\phi_\xi - \frac{\delta W(\chi,\xi)}{\delta \xi}\phi_\chi\right]
$$

Calculating the Dirac brackets we get, as the quantization condition,
that the commutation relation
\begin{equation}
[\xi(x',z), \chi(x,z)] = \frac{i \hbar }{f_0^2}\delta(x - x') \label{quantization}
 \end{equation}
 should be imposed. However, the paraxial quantization condition (\ref{quantization}) hods for the operators acting at the same propagation distance $z$ rather than at the same time as is usually done.

The question of how the "paraxially" quantized operators relate to the usual photons is in order. The issue of paraxial quantization was addressed in Ref. \onlinecite{AML10}. It was shown that the positive frequency part of the electric field operator (Coulomb gauge) in the paraxial approximation has the form
\begin{widetext}
\begin{equation}\label{quantEM}
\hat{E}=i\int_0^\infty d\omega\sqrt{\frac{\hbar\omega}{4\pi\epsilon_0c}} e^{i\omega(t - \frac{z}{c})}\sum_{\mu,m,n} \hat{a}_{\mu,m,n}(\omega) \left(\hat{x}_{\mu} + i\hat{z}\frac{{\bf k}_{\perp}^2c}{\omega} \hat{x}_{\mu} \mathbf{\nabla}_{\perp}\right)\psi_{\mu,m,n}({\bf x},z;\omega)
\end{equation}
\end{widetext}
where $\hat{x}_{\mu}$ are the unit polarization vectors, $\mu=1,2$, ${\bf x} = x \hat{x} + y \hat{y}$ is normal to the propagation direction, ${\bf k}_{\perp} = k_x\hat{x}+k_y\hat{y}$ is the transverse part of the wave vector, $\mathbf{\nabla}_{\perp}=\hat{x}\partial_x+\hat{y}\partial_y$, $\psi_{\mu,m,n}({\bf x},z;\omega)$ make a set of orthogonal polynomials
\begin{equation}\label{orthogonality}
\sum_{m,n} \psi_{m,n}({\bf x},z;\omega) \psi^*_{m,n}({\bf x}',z; \omega) = \delta({\bf x} - {\bf x}')
\end{equation}
e.g. Hermite or Laguerre polynomials. $\hat{a}_{\mu,m,n}(\omega)$ and $\hat{a}^\dag_{\mu,m,n}(\omega)$ are annihilation and creation operators of a photon with the spatial mode $m,n$ and polarization $\mu$ satisfying the standard boson commutation relations.

In the current paper we keep only one polarization and neglect the terms $O(k_\perp^2)$. Then Eq. (\ref{quantEM}) becomes
\begin{widetext}
\begin{equation}\label{quantEM-a}
\hat{E}=i\int_0^\infty d\omega \sqrt{\frac{\hbar\omega}{4\pi\epsilon_0c}} e^{i\omega(t-\frac{z}{c})} \sum_{m,n} \hat{a}_{m,n}(\omega)\psi_{m,n}({\bf x},z;\omega) = \int\frac{d{\omega}}{2\pi} \sqrt{\frac{i\omega}{4\pi\epsilon_0c}} \widehat{A}(x,y,z,{\omega}) e^{i(\beta_0z - \omega t)}
\end{equation}
\end{widetext}
Here the operator $\widehat{A}$ represents the classical amplitude $A$ satisfying the NLS equation (\ref{NLS}) (up to the square root factor). Applying the procedure similar to that used in BEC (see, e.g. Ref. \onlinecite{FV71}) we write
$
\widehat{A} = A_0 + \delta \widehat{A}
$
where the classical amplitude $A_0= \langle \widehat{A} \rangle$ is obtained by averaging the operator $\widehat{A}$ over the coherent state describing the field in the stationary laser beam.
\begin{equation}\label{qu-fluctuations}
\delta\widehat{A} =  \sqrt{i\hbar}\sum_{m,n} \hat{b}_{m,n}(\omega)\psi_{m,n}({\bf x},z;\omega)
\end{equation}
is the fluctuation operator. We have introduced here the new photon operators $\hat{b}_{m,n}(\omega) = \hat{a}_{m,n}(\omega) - \langle \hat{a}_{m,n}(\omega) \rangle $ and $\hat{b}^\dag_{m,n}(\omega) = \hat{a}^\dag_{m,n}(\omega) - \langle \hat{a}^\dag_{m,n}(\omega) \rangle $. These new operators correspond to the fluctuations of the electric field. They obviously satisfy the boson commutation relations and $\langle\hat{b}_{m,n}(\omega)\rangle = \langle \hat{b}^\dag_{m,n}(\omega) \rangle =0 $.

As the last step we assume that $A_0 = f_0 e^{-i\varphi_0}$ and write
\begin{equation}\label{qu-fluctuations-a}
\delta \widehat{A} =  f_0 e^{-i\varphi_0} \left[\frac{1}{2}\widehat{\chi}({\bf x},z;\omega) + i \widehat{\xi}({\bf x},z;\omega)\right].
\end{equation}
The two operators in Eq. (\ref{qu-fluctuations-a}) represent fluctuations of the amplitude and the phase, respectively. Since these quantities are real the operators
\begin{widetext}
\begin{equation}\label{quantization-a}
\begin{array}{ccc}
\widehat{\chi}({\bf x},z;\omega)  & = & \displaystyle\frac{\sqrt{i\hbar}}{f_0}\sum_{m,n}  \left[\hat{b}^\dag_{m,n}(\omega)\psi_{m,n}^*({\bf x},z;\omega)e^{-i\varphi_0}  + \hat{b}_{m,n}(\omega)\psi_{m,n}({\bf x},z;\omega) e^{i\varphi_0}\right], \\
\widehat{\xi}({\bf x},z;\omega) & =& \displaystyle\frac{\sqrt{i\hbar}}{2if_0}\sum_{m,n}  \left[\hat{b}^\dag_{m,n}(\omega)\psi_{m,n}^*({\bf x},z;\omega)e^{-i\varphi_0}  - \hat{b}_{m,n}(\omega)\psi_{m,n}({\bf x},z;\omega) e^{i\varphi_0}\right]
\end{array}
\end{equation}
\end{widetext}
are Hermitian. Using the orthogonality condition (\ref{orthogonality}) and bosonic commutation relation for the operators $\hat{b}^\dag_{m,n}(\omega)$ and $\hat{b}_{m,n}(\omega)$ we may readily verify that the commutation relation (\ref{quantization}) holds. Equations (\ref{quantization-a}) connect paraxially quantized quantities (\ref{quantization}) with the photon operators $\hat{b}^\dag_{m,n}(\omega)$ and $\hat{b}_{m,n}(\omega)$.

\section{Concluding remarks}

We discuss here the role that the finite width of the analog event horizon plays in the dynamics of fluctuations and formation of the spectrum of the Hawking radiation. Fluctuations near the GR event horizon, discussed in the recent papers\cite{BP12,CPF14}, are described by an equation for one field, which differs from equations (\ref{linear-1}) and (\ref{linear-2}) for two fields that follow from the NLS equation for the optical analog event horizon. As a result, the fluctuation modes obtained in Section II differ from those obtained in Ref \onlinecite{BP12,CPF14}. Actually it means that the laboratory frame of the analog systems is not fully compatible with the Schwarzschild frame in GR. The transformation (\ref{x-trans}) and (\ref{z-trans}) needed in order to reach better compatibility. Then the fluctuation modes (\ref{incoming}) and (\ref{outgoing}) become really analogous to those obtained in Ref \onlinecite{BP12,CPF14}. This transformation is singular at $|x| \to 0$, therefore the relevant results hold only outside the width of the horizon $|x|>l_r$. Nevertheless, it is sufficient for our analysis of the scattering matrix in Section III.

Calculating the scattering matrix to within the leading order corrections in the nonlinearity length (which determines the width of the horizon) we come to the conclusion that the Hawking radiation is that of a grey body with the emissivity larger than one. This result is quite understandable since the system is stationary but out of equilibrium and there is a permanent source of energy. This conclusion is certainly not specific for the Kerr nonlinear optical systems, discussed here, and can be readily extended to other systems such as, say, Bose-Einstein condensates.

We also show here how the paraxial quantization (commutation relations at the same propagation distance rather than at the same time) is connected with the conventional quantization of the electromagnetic field. As a result, the paraxial operators $\hat{\chi}$ and $\hat{\xi}$ describing the Hawking radiation can be now converted into regular photon operators.

{\bf Acknowledgement} We are grateful to S. Bar-Ad, M. Ornigotti, R. Parentani, N. Pavloff and G. Shlyapnikov for fruitful discussions. The work was supported by the Israeli Science Foundation.


\begin{thebibliography}{99}

\bibitem{U81} W. G. Unruh, Phys. Rev. Lett., {\bf 46}, 1351 (1981).

\bibitem{H75} S. W. Hawking, Commun. Math. Phys., {\bf 43}, 199 (1975).

\bibitem{H76} S. W. Hawking, Phys. Rev., D {\bf 13} , 191   (1976).

\bibitem{JV98} T. A. Jacobson and G.E. Volovik,  Phys. Rev. D {\bf
58}, 064021 (1998).

\bibitem{R00} B. Reznik, Phys. Rev. D, {\bf 62}, 044044 (2000).

\bibitem{G05} S. Giovanazzi, Phys.Rev.Lett., {\bf 94}, 061302
(2005).

\bibitem{BLV03} C. Barcelo, S. Liberati, and M. Visser, Phys. Rev.
A, {\bf 68}, 053613 (2003).

\bibitem{CFRBF08} I. Carusotto, S. Fagnocchi, A. Recati, R.
Balbinot, and A. Fabbri, New J. Phys., {\bf 10}, 103001 (2008).

\bibitem{RPC09} A. Recati, N. Pavloff and I. Carusotto, Phys. Rev.,
A {\bf 80}, 043603 (2009).

\bibitem{NBRB09} P. D. Nation, M. P. Blencowe, A. J. Rimberg, and
E. Buks, Phys. Rev. Lett., {\bf 103}, 087004 (2009).

\bibitem{FFBF10} I. Fouxon, O.V. Farberovich, S. Bar-Ad and V.
Fleurov, Europhys.Lett., {\bf 92}, 14002 (2010).

\bibitem{BCP14} X. Busch, I. Carusotto, R. Parentani, Phys.Rev. A {\bf 89}, 043819 (2014).

\bibitem{PKRHK08} T. G. Philbin, C. Kuklewicz, S. Robertson, S.
Hill, F. K\"onig, and U. Leonhardt, Science, {\bf 319}, 1367 (2008).

\bibitem{BCCGORRSF10} F. Belgiorno, S. L. Cacciatori, M. Clerici,
V. Gorini, G. Ortenzi, L. Rizzi, E. Rubino, V. G. Sala, and D.
Faccio, Phys.Rev.Lett., {\bf 105}, 203901 (2010).


\bibitem{LIBGRZS10} O. Lahav, A. Itah, A. Blumkin, C. Gordon, S.
Rinott, A. Zayats, J. Steinhauer, Phys. Rev. Lett. {\bf 105}, 240401
(2010).

\bibitem{Jeff14} J. Steinhauer, Nature Physics, {\bf 10}, 864–869 (2014).

\bibitem{RMMPL08} G. Rousseaux, C. Mathis, P. Maissa, T. G.
Philbin, and U. Leonhardt, New J. Phys., {\bf 10}, 053015 (2008).

\bibitem{RMMPL10} G. Rousseaux, P. Maissa, C. Mathis, P. Coulet, T.
G. Philbin, and U. Leonhardt, New J. Phys., {\bf 12}, 095018 (2010).

\bibitem{WTPUL11} S. Weinfurtner, E. W. Tedford, M. C. J. Penrice,
W. G. Unruh, G. A. Lawrence, Phys. Rev. Lett., {\bf 106}, 021302
(2011).

\bibitem{EFB12} M. Elazar, V. Fleurov, and S. Bar-Ad, Phys. Rev. A {\bf 86}, 063821 (2012).

\bibitem{EBFS13} M. Elazar, S. Bar-Ad, V. Fleurov, R. Schilling, An all-optical event horizon in an optical analogue of a Laval nozzle -in Analogue Gravity Phenomenology Lecture Notes in Physics Volume 870, 2013, pp. 275-296 - Springer

\bibitem{R12} S.J. Robertson, J.Phys. B: Mol.Opt.Phys, {\bf 45}, 163001 (2012).

\bibitem{BF14} R. Balbino and A. Fabri, Advances in High Energy Physics, {\bf 2014}, 1 (2014).

\bibitem{C13} I.Carusotto, Proc. R. Soc. A. 470, 20140320 (2014).

\bibitem{J91} T. Jacobson, Phys. Rev. D {\bf 44}, 1731 (1991).

\bibitem{J93} T. Jacobson, Phys. Rev.D {\bf 48}, 728 (1993).

\bibitem{U95} W. G. Unruh, Phys. Rev. D {\bf 51}, 2827 (1995).

\bibitem{C97} S. Corley, Phys. Rev. D {\bf 55}, 6155 (1997).

\bibitem{MP09a} J. Macher and R. Parentani, Phys. Rev. D {\bf 79}, 124008 (2009).

\bibitem{MP09b} J. Macher and R. Parentani, Phys. Rev. A {\bf 80}, 043601 (2009).

\bibitem{BMPS95} R. Brout, S. Massar, R. Parentani, and Ph. Spindel, Phys. Rev. D {\bf 52}, 4559 (1995).

\bibitem{CJ96} S. Corley and T. Jacobson, Phys. Rev. D {\bf 54}, 1568 (1996).

\bibitem{C98} S. Corley, Phys. Rev. D {\bf 57}, 6280 (1998).

\bibitem{CJ99} S. Corley and T. Jacobson, Phys. Rev. D {\bf 59}, 124011 (1999).

\bibitem{LKO03a} U. Leonhardt, T. Kiss, and P. \"Ohberg, Journal of Optics B: Quantum and Semiclass. Opt. {\bf 5}, S42, (2003).

\bibitem{LKO03b} U. Leonhardt, T. Kiss, and P. \"Ohberg, Phys. Rev.
A {\bf 67}, 033602 (2003).

\bibitem{US05} W. G. Unruh and R. Sch\"utzhold, Phys. Rev. D {\bf 71}, 024028 (2005).

\bibitem{SU08} R. Sch\"utzhold and W. G. Unruh, Phys. Rev. D {\bf 78}, 041504 (R) (2008).

\bibitem{FP11} S. Finazzi and R. Parentani, Phys. Rev. D {\bf 83}, 084010 (2011).

\bibitem{BLV05} C. Barcelo, S. Liberati, and M. Visser, Living Rev. Relativ. {\bf 8}, 12 (2005).

\bibitem{FS12} V. Fleurov and R. Schilling, Phys. Rev. A {\bf 85}, 045602 (2012).

\bibitem{CP14} A. Coutant and R. Parentani, arXiv:1402.2514

\bibitem{DY14} A. Davidson, B. Yelin, Phys. Lett. B {\bf 736}, 267 (2014).

\bibitem {LPK12} P. $\acute{\rm E}$. Larr$\acute{\rm e}$, N. Pavloff, and A. M. Kamchatnov, Phys. Rev. B {\bf 86}, 165304 (2012).

\bibitem {LRCP12} P. $\acute{\rm E}$. Larr$\acute{\rm e}$, A. Recati, I. Carusotto and N. Pavloff, Phys. Rev. B {\bf 85}, 013621 (2012).

\bibitem{DG91} I.H. Deutsch and J.C. Garrison, Phys. Rev., {\bf 43}, 3498 (1991)

\bibitem{AW05} A.Aiello and J.P. Woerdman, Phys. Rev. {\bf72}, 060101(R) (2005)

\bibitem{AML10} A. Aiello, C. Marquardt and G. Leuchs,  Phys. Rev. A {\bf 81}, 053838 (2010).

\bibitem{CPF14}   A. Coutant and R. Parentani, S. Finazzizi arXiv:1108.1821v3 (2014)

\bibitem{BP12}   X. Busch and R. Parentani, arXiv:1207.5961v2 (2012)

\bibitem{DR76} T. Damour, and R. Ruffini, Phys. Rev. D {\bf 14}, 332 (1976).

\bibitem{DL08} T. Damour and M. Lilley, arXiv:0802.4169v1 (2008)


\bibitem{FS11} V. Fleurov and R. Schilling, arXiv:1105.0799.

\bibitem{D64} P.A.M. Dirac, Lectures on Quantum Mechanics, Dover Publications, (1964).

\bibitem{FV71} A. L. Fetter und J. D. Valecka, Quantum theory of many-particle systems, McGraw-Hill, (1971).


\end{thebibliography}
\end{document}